# The Cosmology of David Bohm: Scientific and Theological Significance

Richard de Grijs[1,2] and Doru Costache[3,4]

[1] *School of Mathematical and Physical Sciences, Macquarie University, Balaclava Road, Sydney, NSW 2109, Australia;*
[2] *Research Centre for Astronomy, Astrophysics and Astrophotonics, Macquarie University, Balaclava Road, Sydney, NSW 2109, Australia;*
[3] *Sydney College of Divinity's Graduate Research School, Macquarie Park, NSW 2113, Australia;*
[4] *Studies in Religion, School of Humanities, the University of Sydney, Camperdown, NSW 2006, Australia.*

Corresponding author: Doru Costache dcostache@stcyrils.edu.au

**Abstract**: We discuss David Bohm's dual contributions as a physicist and thinker. First, de Grijs introduces Bohm's universe, with an emphasis on the physical quest that led Bohm to the elaboration of an original cosmology at the nexus of science and philosophy. Next, Costache takes his cue from de Grijs' explorations by highlighting the affinity between Bohm's scientific cosmology and patristic ideas that are central to the Orthodox worldview. It is our hope that this approach will stir the interest of Bohm scholars in the Orthodox worldview and also lead Orthodox theologians to nurture an appreciation for Bohm's cosmology.

This article is the outcome of our research within the framework of the project "Science and Orthodoxy around the World" (SOW), running at the Institute of Historical Research of the National Hellenic Research Foundation, Athens (2020–2023).[1] We have chosen to examine David Bohm's thinking for various reasons, beyond our personal interest in his ideas. Since he showed an aptitude for bridging scientific, spiritual, and theological representations of reality, we wondered whether his approach could be extrapolated for the purposes of making sense of the Orthodox worldview in contemporary terms. The first part of this study shows that Bohm himself demonstrated that certain Eastern religious and spiritual philosophies can inspire the scientific quest, on the one hand, and facilitate the presentation of scientific ideas in meaningful ways, relevant to the human experience in its entirety and complexity, on the other hand. Could not the Orthodox worldview be used as a similar platform? We believe that it can. To that end, Orthodox theologians themselves should overcome their scientific apathy and boldly formulate what Dumitru Stăniloae called "a theology of the world" that "reconciles the cosmic vision of the Fathers with a vision which grows out ... of the natural sciences."[2] Without such an effort, tomorrow's David Bohms will never be able to appreciate the Orthodox worldview.

Herein, we focus upon the possibility of articulating holistic and dynamic worldviews—including the theory of everything (after Bohm's fashion) and its ramifications—as a way of bridging the Orthodox representation of reality and contemporary cosmological ideas. After mapping, in the first part of this article, Bohm's scientific contributions and philosophical cosmology, we examine the Orthodox worldview in conversation with his views. Our aim is dual, namely, to determine whether Orthodox theology is compatible with modern cosmological ideas, and to do so from the vantage point of Bohm's holistic and interdisciplinary approach to reality. In turn, the latter will bring to the fore the theological significance of Bohm's cosmological thinking.

## 1. A Theory of Everything[3]

Particle physics and cosmology are the most fundamental sciences that aim at answering humanity's existential questions, *Who are we? Where do we come from?* and *What is the nature of reality?* Of all disciplines in the physical sciences, cosmology is perhaps best placed to make significant inroads towards addressing the complexity governing the universe at large, as it resides at the nexus of physics, philosophy, and even consciousness, offering some insights into crossover aspects of science and spirituality.

Among the great twentieth-century thought leaders in this area, David Joseph Bohm (1917–1992) is perhaps the most poorly known.[4] A student of the "father of the atomic bomb" and leader of the Manhattan Project, Robert Oppenheimer, and a close friend and collaborator of Albert Einstein, Bohm realised that for science to truly benefit mankind, "something" beyond science itself was required. This is the premise of the first half of this essay.

Physicists have long attempted to describe the natural world we can see and explore with our senses using a single, all-encompassing theory, the Theory of Everything.[5] It would combine the macroscopic cosmos described by Einstein's Theory of General Relativity with the subatomic world of Werner Heisenberg's and Niels Bohr's quantum physics and quantum mechanics. Such an all-encompassing theory describing the nature of our physical reality has thus far eluded all efforts. At the basis of this prevailing stalemate is the inherent tension between Einstein and Bohr, exhibited through the fundamental incompatibility of their two sets of equations. Einstein and Bohr were fundamentally unable to agree on the ultimate question: *What constitutes the true nature of reality?*[6]

Bohm's early ideas related to this conundrum gravitated around the commonly held notion that the vacuum among the stars is inert and empty. He posited that the vacuum represents a *plenum*,[7] a space completely filled with matter and energy. The vacuum would thus be infinitely full rather than infinitely empty. Here, *space* refers to a multidimensional concept, a highly varied "structure–process" of which our observable space is merely a projection.[8] In this context, *matter* is represented by small holes in the plenum, which would hence make space a "living organism."[9] Of course, this discussion was not without antecedents. The representation of space as either full or empty has been a source of tension since Antiquity, and Bohm was certainly aware of this.[10] The ancient Greek school of the pre-Socratic philosophers Parmenides (fl. late 6th/early 5th century BCE) and Zeno (ca. 495–ca. 430 BCE) of Elea promoted the idea of space as a plenum. Democritus (ca. 460–ca. 370 BCE) and his mentor Leucippus (fl. 5th century BCE), on the other hand, suggested that space was empty (the "void") in which particles of matter (atoms) can move around freely.[11]

By ascertaining the plenitude of space and by redefining matter against this backdrop, Bohm challenged the basic ideas at the core of the prevailing, classical theories, calling them out for the unclear nature and inherent contradictions of their underlying assumptions. He suggested that physics required a radically new order.[12]

## 2. Undivided Wholeness

Instead of focusing on the contradictions inherent to quantum mechanics on the one hand and general relativity on the other, Bohm appealed to the scientific community to consider what both theories have in common, a property he referred to as "undivided wholeness."[13] This formed the basis of his developing intuition that science needed a new means to connect the very large to the very small, the universe on the largest scales to the subatomic nanoworld. The visible world we see around us—the "explicate order"—is merely a surface order, which is neither deep nor profound; external order is merely skin-deep. In turn, the profound underlying order that connects it all Bohm called the "implicate order."[14] He would spend most of the rest of his career trying to uncover this fundamental, deep order.

Bohm's new way of thinking about the physical universe at any scale, where everything is internally related to everything else and each part of the cosmos contains the whole universe, represented a radically new approach to reality.[15] In this reality, the implicate order is a structure, a constant unfolding and enfolding (embracing) into the classical world, where the explicate order unfolds from the implicate order into the reality as we perceive it.[16] In turn, these insights imply that we need to develop a more organic perception of reality than our prevailing, rigid three-dimensional idea of space where time progresses linearly. Rather than adopting a mechanical worldview, Bohm suggested that by drawing instead on biological processes we might be better off describing the physical world on the basis of more organic processes.[17]

Bohm appears to have wholeheartedly embraced the philosophical implications of his stream of consciousness. He saw direct parallels between the movement of electrons and the possibility of individual human freedom. He compared these grand concepts to what a number of Oppenheimer's other students called the "Russian Experiment." The latter refers to the idea that a Marxist society as a whole would facilitate the transformation of the individual and potentially improve human welfare at large.

Taking these philosophical insights to their natural conclusion led Bohm, in collaboration with his physicist colleague and friend David Pines, to his quantum-mechanical theory of plasma oscillations in a free-electron gas (and, equivalently, in metals; see below for the philosophical equivalence).[18] The movement of electrons in a plasma, a gas-like substance of ionised (charged) particles, appeared to follow the same rules as individuals in the collective defined and constrained by the "Russian Experiment." Bohm realised that, perhaps paradoxically, electrons in a metal, inherently composed of atoms in a lattice structure, were relatively free, similar to individuals in the collective. In turn, his scientific efforts which culminated in the theory now known as "Bohm diffusion"[19]—the diffusion of electrons in a plasma across magnetic field lines (with significant losses, scaling linearly with temperature and inversely with the strength of the magnetic field)—raised the philosophical question as to the extent to which individuals who are members of a collective can have individual freedom.

Bohm's insights made the leading scientists of the day sit up and pay attention. As a case in point, Einstein sent him a letter stating, "I am very astonished about your announcement to establish some connection between the formalism of quantum theory and relativistic field theory. I must confess that I am not able to guess how such unification could be achieved."[20] The same intention, of course, had preoccupied Einstein himself in the last thirty-odd years of his life, although Einstein's efforts ultimately proved fruitless. However, in Bohm's case the quest for a unified theory—in contrast to Einstein's field equations—necessitated consideration of the wild world of quantum physics.[21]

But Bohm's focus on standard quantum mechanics did not last, in essence because of his conversations with Einstein. He was asked to teach a course on the subject at Princeton, and although he did so adequately, he confessed to not fully understanding the subject. And so he set out to write[22] a comprehensive and much lauded monograph on quantum mechanics, titled *Quantum Theory*.[23] Here, unexpectedly, Bohr's theoretical digressions led him to pursue the idea of wholeness, where everything is interconnected at a deeper level of reality.

## 3. Hidden Variables

Yet, Bohm felt that Bohr's classical formalism did not represent a full description of the quantum world. He questioned the orthodox, Copenhagen interpretation of quantum physics espoused by Bohr, Heisenberg, and their followers, for instance the notion that the particle–wave duality of light cannot be broken. Instead, he proceeded to develop his own formalism, which has since become known as Bohm's "hidden variables," perhaps his most enduring intellectual legacy. Resonating with Einstein's conviction that an objective reality and order must underlie the chaos of the unpredictability that characterises quantum physics,[24] but from a very different vantage point, he affirmed that the behaviours of subatomic particles are not chance processes, since the motions of electrons are guided by underlying "pilot waves."[25]

Bohm suspected that an unknown underlying reality, process, or potential somehow informed the behaviour of electrons, for instance in Young's famous two-slit experiment. In essence, he agreed with the view of the Copenhagen school that the quantum world is radically different from our classical view of physics. He believed that the separate events we observe in our three-dimensional world of space and time are, in fact, connected at a deeper level in the quantum world, since they are part of a single system where that separation does not exist. In so doing, he hinted, however, at a degree of unity and order of the quantum world that went beyond the received views of the Copenhagen school. More specifically, among the "hidden variables" Bohm introduced hypothetical, unobservable entities to explain quantum-mechanical properties in the context of a deterministic theory. He objected to the classical quantum-mechanical idea of Heisenberg's indeterminacy between a particle's speed and location, for instance. Einstein fundamentally disagreed, calling Bohm's theory "a physical fairy tale for children."[26] This is surprising, given that Einstein famously declared that "I am convinced God does not play dice."[27] In fact, Bohm's reference to order at the quantum level should have pleased him. But Einstein's opposition is understandable. It is well known that he believed that quantum mechanics is an incomplete description of reality.[28]

Leading physicists, most notably his former mentor Oppenheimer, tried their best to find flaws in Bohm's theory. They were unable to, and so they ignored him. In Oppenheimer's own words, expressed at a conference in Princeton dedicated to Bohm's work, "If we cannot disprove Bohm, then we must agree to ignore him."[29] Wolfgang Pauli, in turn, considered Bohm's "extra wave-mechanical predictions … still a cheque, which cannot be cashed,"[30] while Bohr called him "very foolish."[31] That said, in time Bohm's scientific contributions received due recognition, as Robert Russell has already shown[32]—and so did his philosophical thinking. Specifically, Bohm's theoretical framework is not subject to the usual measurement problems. Once the latter important constraint is properly recognised, his advances are (finally) set to receive their due recognition.

## 4. Krishnamurti's Influence

Bohm's philosophical leanings became more overt during the next stage of his career, especially under the influence of Jiddu Krishnamurti's ideas about the observer and the observed.[33] He realised that Krishnamurti's Eastern philosophy and his own physical framework appeared to have much in common, which in its simplest form can be expressed as "the observer is the observed."[34] Specifically, Krishnamurti's ideas suggested that the act of observing something inwardly—an emotion, an attitude, a thought—changes it, which resembled the basic tenets of quantum physics.

Against this backdrop, Bohm naturally reverted back to his earlier work on the big picture in physics. Leaving his hidden variables behind, he once again focused on the challenging situation, with physicists, foremost Einstein, having been unable to reconcile quantum theory with general relativity. He was most concerned that modern physics appeared to require two foundational theories rather than a single, overarching framework. In turn, this spawned his thinking of the implicate order. But his mind was grappling with even bigger questions—regarding humankind's wellbeing—at the nexus of science, philosophy, and religion.[35]

Bohm's and Krishnamurti's ideas met in their respective descriptions of the nature of thought, reality, and consciousness, considered a coherent whole that is never static nor complete, but a continuous "holomovement" between the enfolding and the unfolding aspects of reality.[36] Both Bohm and Krishnamurti questioned the foundations of their own areas of emphasis, ultimately rejecting the prevailing orthodox definitions of physics, or of science more generally, and of society, as very limiting. Krishnamurti's fundamental tenet was that thought itself is an actual movement in our lives, a physical and physiological movement of great power. Bohm was profoundly influenced by Krishnamurti's Eastern philosophy. His attraction to Krishnamurti's philosophy may be best described as his adherence to the wholeness of life. Life, nature, and consciousness were a single, indivisible whole, a notion shared by both men.[37]

Krishnamurti's ideas, at their essence, address the separation between the observer and the observed in a similar way as quantum physics defines these concepts as forming part of a whole—no wonder Bohm's interest in his views. It also was Bohm's conviction that one has to transcend science that brought him within Krishnamurti's philosophical sphere.

## 5. A Split Duality

Bohm next developed a new theory, which he referred to as a "structure–process." The underlying idea is that the basis of our reality is part of a process.[38] More precisely, reality does not emerge from particles moving in space-time, but from a process from which both particles and space-time can emerge. Unexpectedly, this new approach to reality facilitated the resurrection of Bohm's earlier notion of hidden variables. His insights revealed that Young's double-slit experiment was naturally consistent with his own theory of hidden variables. The split duality of electromagnetic radiation, manifested by its dual form of waves and particles, could now be explained—although this notion was strictly forbidden in the context of the standard quantum theory. In essence, Bohm's theory suggested that particle trajectories explain the interference pattern resulting from Young's double-slit experiment.[39]

The resulting framework, known as the quantum potential, implied the interconnectedness of the universe at the fundamental quantum level. A "dip" in the quantum potential corresponds to a rate of change, which is equivalent to a force. In turn, underlying variations in the quantum potential give rise to the observed interference pattern. The quantum potential energy only functions on those occasions when quantum phenomena are actually of importance. In other words, when a particle approaches the double slits, instead of adopting the classical physics waveform, the quantum potential is responsible for organising the particle's trajectory. This theory is now known as the pilot-wave or the Bohm–de Broglie theory.[40]

## 6. Wholeness

Remarkably, Bohm's hidden variables theory was in agreement with this new development, in particular with predictions of the nonlocality of reality, that is, the interconnectedness of the universe, where Bohm's hidden parameters would be nonlocal.[41] This concept is problematic from a classical perspective, where communication is limited by the finite speed of light, suggesting that events are localised rather than nonlocal, and information is not transmitted instantaneously to any remote corners of the universe. The universal acceptance among physicists of the finite speed of light is, hence, a major obstacle to Bohm's concept of hidden variables as a nonlocal concept.

The theory behind the nonlocality of reality also suggests that there exists a hidden regime of reality that cannot be accessed and which will, hence, remain beyond the realm of science and the scientific method[42]—a domain of reality that, perhaps, we have not yet understood, or at least not well enough. If the putative quantum potential is eventually confirmed, the existence of nonlocality, the profound interconnectedness of the physical universe, is consequently also confirmed. What matters for the present discussion is that the premise of wholeness pertaining to the quantum potential framework also agrees with Bohr's classical quantum physics ideas.

Bohr posited that when dealing with quantum-mechanical reality, one cannot follow a classical physics approach of creating silos for the analysis of individual ideas. However, one could mathematically unfold individual elements of space into the whole, subsequently enfolding them into their own domains once the analysis has been completed. This would work best if done cyclically. Therefore, what may appear like a continuous trajectory is, in fact, a series of unfoldings and enfoldings.[43] Of course, all this needs validation, especially through observation and experiment. Validation of Bohm's ideas regarding the wholeness of the universe, of its interconnectedness at the most fundamental level, would then directly support the unification and holism of all aspects of the universe. In turn, this would provide, it seems, a direct parallel to religious ideas that have long advocated a similar description of the universe, both seen and unseen.

At a fundamental level, Bohm's theory suggests that out of perceived emptiness, resembling the "vacuum state," particles interact with, respond to, and are informed by a quantum potential which allows the cosmos to emerge. This information in the quantum potential makes it possible for the universe as we know it to exist. In other words, everything we know (and do not know—yet) is equivalent to information, which at some point is expected to unfold into reality. The implicate is waiting to become explicate, thus showing the true nature of the split duality of particles and trajectories.

Reality is thus undivided wholeness, combining life, the universe, and everything.

However, Bohm remained undecided as to whether a higher intelligence, a "God" if you like, was present in his implicate order. He continued thoughtfully, "The implicate order does not rule out God, nor does it say there is a God. But it would suggest that there is a creative intelligence underlying the whole, which might have as one of the essentials that which was meant by the word 'God'."[44]

Similar views have been put forward with increasing frequency in more recent years, corroborating Bohm's integrative approach to philosophy, science, and spirituality.[45] And while Bohm's approach has already received attention within the Western theological context,[46] Orthodox theologians are yet to assess its relevance. The ensuing reflections propose a way of doing so.

## 7. Theological Narratives of Everything[47]

As de Grijs showed above, Bohm's thinking draws parallels to religious ideas, bridging science, philosophy, spirituality, etc. As such, it diverges from the abstract rationalisations and compartmentalised knowledge of modern culture.[48] Bohm's thinking therefore deserves attention from theologians, including Orthodox ones, and especially from researchers of early Christian and medieval cosmology. Above all, Bohm's ideas fascinate by integrating what Pierre Teilhard de Chardin called "the Whole and the Person," the universe and consciousness. True, for Teilhard it is "the Christian phenomenon" that harmonises these poles of existence.[49] The extent to which Bohm would have sympathised with Teilhard's solution is uncertain, but their respective perceptions are not inconsistent.[50] Bohm himself contemplated the nexus between "the thing and the thought"[51] or "thought and non-thought"[52] within the context of a reality that ultimately remains "unknown and unknowable."[53] But to compare their respective approaches falls outside the scope of this study. What matters is that their efforts converge into the modern quest for comprehensive views of reality, with or without articulating theories of everything. These views of reality, by definition, integrate all things, mind and matter. Even more important is that these views resonate with the related concerns of many early Christian and medieval theologians.

The tradition documents many corresponding attempts. Certain early Christian and medieval authors compiled catalogues of reality's components, listing the objects that populate the created realm. Others produced encompassing depictions of the universe. By contemporary standards, their efforts do not count scientifically. However, as Werner Heisenberg and Michio Kaku observed that some ancient intuitions about reality foreshadow contemporary physics,[54] the same must hold true for the early Christian and medieval worldviews that redeployed the ancient intuitions within other cultural contexts. By drawing on Heraclitus and the insights of other ancient Greek theorists, Bohm himself agreed to Heisenberg's and Kaku's point implicitly.[55] And while he did not seem aware of the efforts of the early Christian and medieval theologians to map reality, these belong with the broader history of scientific ideas.[56] Therefore, they should be treated after the manner of Bohm's own approach to Eastern religious philosophies. Here are two examples from the Christian tradition that would require this kind of treatment.

An anonymous second-century treatise known as *Letter to Diognetus* includes a list of cosmic regions. These are "the skies and things celestial, the earth and the earthly things, the sea and things aquatic, (as well as) fire, air, the abyss, (namely,) things on high, things in the depths, things in between."[57]

While mapping the cosmos, the passage refers to the fundamental elements known at the time, earth, water, air, and fire. These are the province of contemporary quantum physics. But the small scale of fundamental elements—"the depths"—is but one side of things. Reality also encompasses the familiar universe, the sky and the earth, and all things that populate them. In a brushstroke, the passage links the two dimensions of reality we currently access through Bohr's quantum physics and Einstein's relativity. But the above catalogue is not complete. Given the theological nature of the treatise, "the depths" and the universe feature together as a dynamic field where divine providence operates, God's Logos "organising, defining, and connecting all things."[58] The divine activity links the various parts of reality, securing order much in the way Bohm's "creative intelligence," mentioned at the end of the previous section, does. It does so from within and through the laws of nature, not superimposing order from outside (the verb *hypotetaktai*, translated above by "connecting," suggests just that, the preposition *hypo* denoting a process that occurs within things).

The idea echoes Heraclitus, for whom the wise person is able to grasp "the will that leads all things through all things."[59] And since modern physicists, including Bohm, credit Heraclitus with scientific intuitions about reality, the same could be said about later, Christian authors who shared in his views.

Apart from the explicitly theological statements of *Diognetus*, Bohm would have been thrilled to read these lines. After all, he searched for ways of explaining how the quantum world and the relativistic universe—the implicate and the explicate, the enfolding and the unfolding orders—are woven together.[60] Immediately relevant, here, is that *Diognetus* bridges the various layers of the ancient worldview into a comprehensive picture.[61]

*Diognetus* completes the cartography of the universe by referring to what our culture knows as the anthropic cosmological principle: "God loves human beings. He created the cosmos for them and to them he subjected all things on the earth."[62] This is not a detailed point about the mutual attunement of humankind and the cosmos, but it definitely shows that a holistic cosmology cannot ignore human presence, or mind, or consciousness. The author, therefore, was not insensitive to the concerns more recently expressed by Teilhard and Bohm.[63]

Centuries later, Maximus the Confessor (d. 662) presented a much more detailed map of reality. In his *Book of Difficulties*, he proposes a genuine theory or narrative of everything.[64] The relevant chapter outlines five layers of reality, each containing two poles: first, the created and the uncreated; second, the intelligible and the sensible orders; third, the sky and the earth; fourth, paradise and the inhabited land; fifth, male and female. Every second term of each layer shelters a narrower layer within itself—a smaller circle within the larger one.[65] The largest circle is the ultimate division of reality, of God and the universe. The narrowest layer is the gender division, comprising masculinity and femininity. The other three circles contain the visible and invisible universe in its entirety, then only the visible universe, divided into the sky and the earth, and finally the earthly domain, divided into paradise and the inhabited land. This schematic representation does not exhaust reality—the terrestrial biosphere is not even mentioned—but it still sketches a comprehensive worldview. Furthermore, it combines a range of cultural and disciplinary perspectives, that is, doctrinal (the first circle), metaphysical (the second circle, of Platonic resonance), scientific (the third circle, of Aristotelian resonance), and scriptural (the final two circles).[66] But more important are two further details of Maximus' theory. On the one hand, the regions of the created realm and all populations nestled within them, human and nonhuman, can be unified into a higher form of complex harmony. On the other hand, the agent of unification is the human being, mind and body alike—as long as it adopts a divine, or virtuous, lifestyle that leads to insight.[67]

What makes possible the unification of all layers of created reality is the fact that within each particular being are found—enfolded, Bohm would say—the characteristics of its species. In like manner, on a deeper level, within each species enfolds the nature of the universe. In Bohm's words, "in some sense each region contains a total structure 'enfolded' within it."[68] Since the universe's stuff is the nature of all things, all things are one, they are compatible, they are present in one another, naturally intersecting and combining in more complex structures within the universe.[69] No wonder the human being, related to all things—"constituting something like a natural connector of the whole through mediating by its own parts between the extremities"[70]—is able to access them and to contribute to their unity. The observer and the observed are one, Bohm would say.[71] Here, Maximus and the tradition he represents come close to the conclusions of Ted Peters, Carl Peterson, and Russell regarding the need for "fertile" theological models, such as Bohm's own, able "to progressively open up new vistas and expand horizons of our understanding."[72]

Both theologians, the author of *Diognetus* and Maximus, agree that to grasp human existence entails to understand the place humankind has within the universe and the universe in its entirety. This amounts to saying that we need a theory or a narrative of everything—which corresponds to Bohm's fundamental intuition about reality, as discussed in the first half of this study.

**8. Chaos to Cosmos**

The examples reviewed above document an interest in cataloguing realms and beings; they also bring to the fore the dynamic of cosmic unity. As it happens, there are patristic authors whose ideas come even closer to modern concerns, including Bohm's, pertaining to bridging the quantum world and the relativistic universe. Once again, let us consider two examples.

In *An Apology for the Hexaemeron*, Gregory of Nyssa (d. ca. 395)[73] discusses the processes that led from chaos to the universe. He addresses this topic through his theory of matter and by pondering how the unified light of the origins has become the many lights of the cosmos. He takes as a starting point the view that "some wise and organising principle lies within each of the beings."[74] This matches Bohm's idea of God as "creative intelligence," mentioned above, as much as *Diognetus* does. This organising principle in nature facilitates the generation of beings by activating the fundamental elements and qualities towards material combination.[75] Gregory's particle physics is not important. What matters is his view of nature as the outcome of converging qualities, spurred by the divine factor.

Corresponding to *Diognetus*, Gregory's divine principle, the Logos—designated as "wisdom, will, and power"[76]—is not an extraneous factor. It operates from within the universe by facilitating the combination of the elements. This understanding is reminiscent of Heraclitus' point that "all beings become in accordance with the Logos."[77] That said, for Gregory matter emerges from the aggregation of immaterial qualities of which we know but that cannot be grasped before we see them as concrete beings: "none of these [qualities], of itself, constitutes matter; they become matter when they converge into one another."[78] This resonates with our experience with the quantum world via its material phenomena—or with Bohm's implicit, enfolding order through the explicit, unfolded one.

The process of universe-making is analogous to the process of matter-making. As a preliminary stage for discussing the emergence of the cosmos, Gregory considers the manner in which the original darkness at Genesis 1:2 had become the light of Genesis 1:3. The darkness of chaos possessed the potentiality of light, or order, and became light at God's prompting.[79] The divine energy—"radiating through the darkness and flow of nature"[80]—transformed the darkness into light or the chaos into order. To explain how this occurred, Gregory makes reference to Aristotle's idea of actualisation: light is the actualised form of the potentiality represented by darkness. Here, he anticipated Bohm's own notion of unfolding order.[81]

Gregory's physics of light does not end here. In the universe are found various forms of light—solar, lunar (reflected solar), stellar, etc.—not merely one.[82] But how did the unified light of the first day of Genesis 1 become the many lights of the fourth day? In *Apology* 8–9 and 64–74, Gregory tackles this topic epistemologically and ontologically. Epistemologically, the unified light of the first day denotes reality as "perceived by the divine eye."[83] It is God's grasp of things; a theological and a contemplative lens are at play here. In turn, the many lights of the universe, materialised in the seven spheres of Ptolemaic cosmography and in the celestial bodies that populate them, represent the human viewpoint, analytical and scientific.[84] In turn, ontologically, the differentiation of the original light into the universe's many lights denotes physical processes: light "was gathered within itself, coextensive with the whole, but after [the commandment] it diversified into what was shared and what was distinct in regard to its parts."[85] Gregory does not attempt to bridge the epistemological and the ontological explanations, but what matters is that both, and especially the ontological one, affirm the manifestation of the quantum or enfolding order as the unfolding shape of the relativistic universe.

What causes differentiation is cosmic motion, the fact that the various parts of the universe move differently. What conditions their specific movements are the manners in which they combine the qualities inherent to the light—or fire—that is fundamental of all things.[86] Like Heraclitus,[87] Gregory believes that all beings derive from the dynamic factor called fire[88] or, in scriptural parlance, light. Thus, whether or not he knew of Heraclitus' views, Gregory had a strong intuition of the dynamics pertaining to the manifestation of what we call the quantum world into/as the material universe—in Bohm's terms, the unfolding of the implicate into the explicate order. Thus, as he was concerned with the same issue, Bohm would have appreciated this insight.

The second witness is John Damascene (d. 749), who, in *An Exact Exposition of the Orthodox Faith*, pursued a similar idea, although not to the extent to which Gregory did. Chapters 19–24 of the text's critical edition map the universe by showing what cosmic regions correspond to the fundamental elements known to the ancients; also, what natural phenomena occur and what kinds of beings are found in those places. Alongside the four material elements John lists a more profound level of reality, the sky, which shelters all of creation: "the sky is the container of the visible and the invisible creations."[89] The cosmic sense of our own age is already present here, in a nutshell. And although in the next line John mentions invisible creations—namely, the noetic beings called angels—his worldview is basically physicalist. The sky is neither the metaphysical realm of the eternal forms nor the religious heaven. It is of one piece with the four material elements of earth, air, fire, and water.

God created the sky—the space or the container of everything—and the four elements out of nothing. It is from these elements, then, that God made all else, "animals, plants, and seeds."[90] There is no ontological gap within the created universe. There is an unbroken continuity between the fundamental elements and the world of our experience, including the earthly ecosphere.

John could not explain how the fundamental elements combine in order to engender the realities of our experience. Nor is he eager to engage critically the scientific knowledge of his time. He reviews several hypotheses about the nature of the sky, but does not choose between them.[91] His reluctance matches his conviction that the sky's nature eludes us,[92] a sentiment Bohm shared, as we have seen in the first part of this study. As we know, the nature of the universe still eludes us, with more than ninety percent of it being composed of exotic layers of dark matter and dark energy. Nevertheless, John shows appreciation for other aspects of the available physics. He believes that all things around us are made of the fundamental elements God created out of nothing. It is at this juncture that he depicts the various layers of reality together with the fundamental elements. Thus, in chapters 21–24 he describes the qualities of the four elements and then explores the beings and the phenomena that derive from them.

As its heading announces, chapter 21 deals with "light, fire, and the luminaries, that is the sun, the moon, and the stars." Many other astronomical and astrological topics make cameo appearances. His discourse about fire includes the primordial light and the celestial bodies; the orbits of the seven "wanderers" or planets, so called because they move contrary to the apparent orbits of the "fixed" stars; the solstices and the seasons; the zodiacal signs, comets, and eclipses; the solar and the lunar years, and the phases of the moon. As with the sky, he says, we ignore the true nature of these beings and phenomena.[93] John is aware of the limitations of the available science and, it seems, longs for more precise explanations of the cosmos. He anticipates Bohm's own attitude, outlined in the first half of this study.

John approaches the other elements in the same fashion, discussing in chapter 22 the nature of air and the atmosphere; in chapter 23 the qualities of water and the forms it takes, seen as the habitat of aquatic species; and in chapter 24 the fundamental element of earth and the Earth as a place of human habitation, shared with animals, birds, and plants.[94]

These descriptions appear to a modern reader either as overly simplistic or as downright erroneous. It is for this reason that there is no point in presenting them in any detail. But what matters is that John was able to establish, within the limitations of the available sciences, a relation between the fundamental elements and the cosmic regions, all of which he ultimately understood as belonging together within the one expanse of the sky, the universe. What secures the link between all things is the fact of being created. This explanation anticipates the criterion of "explanatory adequacy" formulated by Peters and Peterson.[95] It cannot satisfy a modern reader, including Bohm, but it is a way of affirming the homogeneity of the universe and the connection between fundamental physics and cosmology, or the implicate and the explicate orders. After all, as de Grijs observes, for Bohm "everything is internally related to everything else and each part of the cosmos contains the whole universe."

## 9. Further Patristic Topics

Bohm would, perhaps, have been more interested in other early Christian ideas, like the microcosm,[96] inherited from the classical culture and developed further. We have seen above that many early Christian authors contemplated the ontological solidarity between humankind and the cosmos, but some preferred to do so through the lens of the microcosm.

For example, Gregory the Theologian (d. ca. 389) shows that human nature is a composition or a mixture of qualities and processes encountered throughout the universe.[97] As such, the human being is a "world in miniature," a microcosm.[98] Elsewhere, he presents human nature as the midpoint between the intelligible (accessible through noetic perceptions) and the sensible (accessible through the senses) regions of the cosmos. Whereas before humankind's emergence these regions "remained within their specific boundaries" and silently sang different praises,[99] the Logos decided to make the human being "a kind of second world, great within the small one, another angel, a composite worshipper."[100] The term *microcosm* does not appear, but the sense that the "composite worshipper" gathers within its nature the elements of the visible and the invisible universe is inescapable. Furthermore, the passage strikingly presents the "second world," the human being, as "great within the small one," therefore larger than the universe. This perspective is provocative, and while

Gregory is not interested in detailing his suggestion, it took the genius of Maximus to explain, centuries later, that the human microcosm has the calling to change the universe into a "large human being" through the successive transformative unifications sketched above.[101] To paraphrase Bohm, this amounts to saying that "the thing" becomes "thought," which would not be a strange avenue for him to contemplate. After all, he wondered, "May not thought itself thus be a part of reality as a whole?"[102]

Bohm would be content with other patristic ideas, too. Perhaps one of his most interesting contributions is his proposal regarding the *rheomode*, the "flowing mode" of thinking and talking about the universe's fluid nature.[103] His concept of an "Undivided Wholeness in Flowing Movement" is crucial for clarifying how mind and matter interact, but also how the implicate order of the enfolded becomes the explicate order of the unfolded. However, we shall not consider either his perception of the nature's constant flow or his point that in order to capture this situation we need a language of verbs and adverbs, not nouns and adjectives. In turn, it is more important to show that certain early Christian and medieval theologians had a similar grasp of the flowing nature of reality. Athanasius of Alexandria (d. 373), for instance, presents the universe as fundamentally "flowing, insubstantial, and mortal" or "fluid and dissolvable."[104] And even though he neither developed Bohm's verbal and adverbial language nor appraised the situation on a global scale, others after him (e.g., Gregory of Nyssa, Maximus the Confessor) developed this intuition, defining nature as movement and describing it in the dynamic vocabulary of Aristotelian actualisation, as we have seen above. This intuition is shared by contemporary Orthodox thinkers.[105] Heraclitus, the philosopher of movement who inspired Bohm so much,[106] would have been proud of his patristic heirs, and so should have been Bohm himself.

In this light, and despite the historical gulfs separating him from the authors discussed above, it is conceivable that Bohm would have been excited about their ideas. Especially their concepts of the microcosm encapsulating and generating the macrocosm, and of the chaos of the fundamental elements transformed into cosmos, share an affinity with his view of implicate order as origin of our universe. Of equal significance are the early Christian and medieval views of the universe as anthropically conditioned—and as possessing an infrastructure that draws on the Logos behind all things—as well as their narratives of everything. These ideas prefigure Bohm's own quest for bridging "the thing and the thought." In particular, the patristic perception of reality—holistic and dynamic—of the chaos morphed into the ordered universe anticipates Bohm's cosmology, where enfolding patterns, operating under the guise of hidden variables, are the source of the unfolding, ordered universe. In this light, Bohm's ideas and the Orthodox worldview appear to be wonderfully compatible. Their similarity deserves further study both as such and for the purposes of articulating an Orthodox "theology of the world" (to paraphrase Stăniloae) able to engage the cosmological ideas of our age.

In turn, Bohm's profound cosmological thinking provides contemporary Orthodox theologians with new opportunities for grasping and for communicating certain traditional intuitions about reality, especially in regard to topics such as the parts and the whole, and matter and consciousness. It remains to be seen how quickly contemporary theologians will turn to Bohm's "postmodern world of tomorrow," as Peters has it,[107] to find answers to their quest for wholeness and purpose,[108] or for making sense of the world we live in and the elusive mechanisms at the heart of things. For now, contemporary Orthodox theologians seem far less able to grapple with the cosmos than their predecessors. Appropriating Bohm's cosmic philosophy might help them to retrieve the early Christian and medieval roots of cosmic theology, as well as a sense of what their tasks might be when it comes to articulating the Christian worldview, today and tomorrow.

*The authors report there are no competing interests to declare.*

Endnotes

[1] This publication has been implemented within the framework of project "Science & Orthodoxy around the World" (SOW), which was made possible through the support of a grant from the Templeton World Charity Foundation, Inc. The opinions expressed in this publication are those of the authors and do not necessarily reflect the views of Project SOW or the Templeton World Charity Foundation, Inc.

[2] Dumitru Stăniloae, *Theology and the Church*, trans. Robert Barringer (Crestwood, NY: St Vladimir's Seminary Press, 1980), 224 (slightly edited).

[3] The first half of this study is de Grijs' contribution.

[4] Olival Freire Jr, *David Bohm: A Life Dedicated to Understanding the Quantum World*, Springer Biographies (New York: Springer, 2019); Robert John Russell, "The Physics of David Bohm and Its Relevance to Philosophy and Theology" *Zygon* 20:2 (1985): 135–58.

[5] Sean Carroll, *From Eternity to Here: The Quest for the Ultimate Theory of Time* (New York: Penguin Books USA, 2010); Stephen W. Hawking, *The Theory of Everything: The Origin and Fate of the Universe* (London: Phoenix Books, 2006); Steven Weinberg, *Dreams of a Final Theory: The Scientist's Search for the Ultimate Laws of Nature* (New York: Knopf Doubleday Publishing Group, 2011).

[6] See Ramin Skibba, "Einstein, Bohr and the War over Quantum Theory," *Nature* 555 (2018): 582–4.

[7] David Bohm, *Wholeness and the Implicate Order* (London: Routledge, 1980), 243; Lee Nichol (ed.), *The essential David Bohm* (London: Routledge, 2003), 150 (Conversations with David Bohm).

[8] Basil J. Hiley, "David Joseph Bohm. 20 December 1917–27 October 1992," *Biographical Memoirs of Fellows of the Royal Society* 43 (1997): 106–31; Lee Nichol (ed.), *The Essential David Bohm* (London: Routledge, 2002).

[9] David Schrum, quoted in the documentary *Infinite Potential: The Life and Ideas of David Bohm* by Paul Howard (2020): 01:30. Full transcript: https://johnmarkmorris.com/2020/07/08/bohmian-mechanics-and-npqg-iii/.

[10] Lee Nichol (ed.) *The essential David Bohm* (London: Routledge, 2003), 152 (Conversations with David Bohm).

[11] W. K. C. Guthrie, *A History of Greek Philosophy*, vol. 2: *The Presocratic Tradition from Parmenides to Democritus* (Cambridge University Press, 1969), 26–67, 88–100, 383–413; Paul T. Keyser and Georgia L. Irby-Massie (eds), *The Encyclopedia of Ancient Natural Scientists: The Greek Tradition and Its Many Heirs* (London and New York: Routledge, 2008), 235–6, 626, 846–7; Bertrand Russell, *A History of Western Philosophy* (London: Simon & Schuster, 1972), 64–5.

[12] David Bohm, *Wholeness and the Implicate Order* (London: Routledge, 1980), Chapter 5.

[13] Bohm, *Wholeness*, Chapter 6.

[14] *Ibid*.

[15] See Russell, "The Physics of David Bohm," 137–41.

[16] Bohm, *Wholeness*.

[17] David Bohm, *Unfolding Meaning* (London: Routledge, 1985); William Seager, "The Philosophical and Scientific Metaphysics of David Bohm," *Entropy* 20 (2018): 493.

[18] David Bohm and David Pines, "A Collective Description of Electron Interactions: I. Magnetic Interactions," *Physical Review* 82:5 (1951): 625–34; idem, "A Collective Description of Electron Interactions: II. Collective vs Individual Particle Aspects of the Interactions," *Physical Review* 85:2 (1952): 338–53; idem, "A Collective Description of Electron Interactions: III. Coulomb Interactions in a Degenerate Electron Gas," *Physical Review* 92:3 (1953): 609–25. David Pines, "A Collective Description of Electron Interactions: IV. Electron Interaction in Metals," *Physical Review* 92:3 (1953): 626–36.

[19] David Bohm, *The Characteristics of Electrical Discharges in Magnetic Fields*, ed. A. Guthrie and R. K. Wakerling (New York: McGraw-Hill, 1949).

[20] Albert Einstein, Letter to David Bohm, 6 December 1951; https://www.newsweek.com/albert-einstein-letters-science-politics-god-auction-625290#slideshow/625294.

[21] Isaacson, *Einstein*, 458.

[22] A. Kojevnikov, "David Bohm" (2021), in *Encyclopedia Britannica*, https://www.britannica.com/biography/David-Bohm.

[23] David Bohm, *Quantum Theory* (New York: Dover, 1951). See also David Bohm in *Infinite Potential: The Life and Ideas of David Bohm* at min. 23:25. Documentary by Paul Howard (2020). Full transcript: https://johnmarkmorris.com/2020/07/08/bohmian-mechanics-and-npqg-iii/.

[24] Isaacson, *Einstein*, 323–36, 352–3, 460–5; David Bohm, "A Suggested Interpretation of the Quantum Theory in Terms of 'Hidden' Variables. II," *Physical Review* 85:2 (1952): 180–93, esp. 189. No wonder his acknowledgement of Einstein at the end of the first part of his study; David Bohm, "A Suggested Interpretation of the Quantum Theory in Terms of 'Hidden' Variables. I," *Physical Review* 85:2 (1952): 166–79, esp. 179.

[25] Bohm, "Variables. II," 191.

[26] Lee Smolin, *Einstein's Unfinished Revolution: The Search for What Lies Beyond the Quantum* (London: Penguin UK, 2019; ebook), 195 n.3; Peat, *Infinite Potential*, 132.

[27] Albert Einstein, Letter to Max Born, 4 December 1926. Albert Einstein Archives, reel 8, item 180. See also *The Collected Papers of Albert Einstein*, col. 15: *The Berlin Years: Writings & Correspondence, June 1925–May 1927* (English transl. suppl.), 403.

[28] Albert Einstein, Boris Podolsky, and Nathan Rosen, "Can Quantum-Mechanical Description of Physical Reality Be Considered Complete?" *Physical Review* 47 (1935): 777–80.

[29] Adam Becker, *What Is Real? The Unfinished Quest for the Meaning of Quantum Physics* (New York: Basic Books, 2018; ebook), 158–9.

[30] Wolfgang Pauli, Letter to Bohm, 3 December 1951. In: *Wolfgang Pauli, Scientific Correspondence*, Vol. IV, Part I, ed. K. von Meyen (Berlin: Springer, 1996), 436–41.

[31] Becker, *What Is Real?* 591.

[32] Russell, "The Physics of David Bohm," 141–8.

[33] Bohm in *Infinite Potential* at min. 37:08.

[34] Jiddu Krishnamurti, *Freedom from the Known*, ed. Mary Lutyens (London: Gollancz, 1969); see also *Collected Works*, vol. 4 (Krishnamurti Foundation America, 2012), available at https://www.kfa.org.

[35] Bohm in *Infinite Potential* at min. 39:27.

[36] Bohm, *Wholeness*, Chapter 6.

[37] Bohm in *Infinite Potential* at min. 45:07.

[38] D. Bohm and B. Hiley, *The Undivided Universe: An Ontological Interpretation of Quantum Theory* (London: Routledge, 1993); B. Hiley, "Stapp, Bohm and the Algebra of Process," *Activitas Nervosa Superior* 61 (2019): 102–7.

[39] Basil J. Hiley, "Bohmian Non-commutative Dynamics: History and New Developments," arXiv:1303.6057 (2013). See also Nichol, *The Essential David Bohm*.

[40] Sheldon Goldstein, "Bohmian Mechanics," in *The Stanford Encyclopedia of Philosophy* (Summer 2021 Edition), Edward N. Zalta (ed.), https://plato.stanford.edu/archives/sum2021/entries/qm-bohm/.

[41] Ibid.

[42] See Russell, "The Physics of David Bohm," 148–9.

[43] *Wholeness*, 204, 211, 233, 230–1, 259–60. See Louwrien Wijers, "Unfolding the Implicate Order: Excerpts from interview with David Bohm," https://www.atisma.com/spiritart/bohm.htm (1989).

[44] See Ted Peters, "David Bohm, Postmodernism, and the Divine," *Zygon* 20:2 (1985): 193–217, esp. 208–212.

[45] See Ted Peters and Carl Peterson, "The Higgs Boson: An Adventure in Critical Realism," *Theology and Science* 11:3 (2013): 185–207, esp. 185–7, 195–8.

[46] See Peters, "David Bohm," and Russell, "The Physics of David Bohm."

[47] The following sections are Costache's contribution.

[48] Peters, "David Bohm," 193 (see ibid. 194–8); Peters and Peterson, "The Higgs Boson," 199–201; Russell, "The Physics of David Bohm," 151.

[49] Pierre Teilhard de Chardin, *Le phénomène humain* (Paris: Seuil, 1956), 331.

[101] See Costache, "Mapping Reality," 381–90.
[102] *Wholeness*, xi.
[103] *Wholeness*, xiv, 14. See Russell, "The Physics of David Bohm," 141.
[104] *Against the Gentiles* 41. See Costache, *Humankind and the Cosmos*, 243–4.
[105] See Doru Costache, "The Orthodox Doctrine of Creation in the Age of Science," *Journal of Orthodox Christian Studies* 2:1 (2019): 43–64; Christopher C. Knight, *Eastern Orthodoxy and the Science-Theology Dialogue* (Cambridge University Press, 2022); Alexei V. Nesteruk, *The Sense of the Universe: Philosophical Explication of Theological Commitment in Modern Cosmology* (Minneapolis: Fortress Press, 2015).
[106] *Wholeness*, 61. Peters, "David Bohm" 198, 204–205.
[107] Peters, "David Bohm," 193. See Peters and Peterson, "The Higgs Boson," 201–3.
[108] Russell, "The Physics of David Bohm," 153–4.

## Biographical Notes

Richard de Grijs is a professor of astrophysics at Macquarie University in Sydney, Australia. He has a keen interest in the history of science. He is an Associate Editor of the *Journal of Astronomical History and Heritage* and Specialty Chief Editor (Fundamental Astronomy) of *Frontiers in Astronomy and Space Sciences*.

Doru Costache is a Romanian Orthodox clergyman living in Australia and an associate professor in patristic studies at the Sydney College of Divinity. He also is an Honorary Research Associate in Studies in Religion at the University of Sydney. He serves as co-editor of *Christian Perspectives on Science and Technology*.